\documentclass[12pt,numerical, amsmath,showpacs, amssymb, reprint,hmargin=1.2in,vmargin=1.2in]{revtex4-1}
\usepackage{hyperref}
\usepackage[utf8]{inputenc}
\usepackage{pgf, tikz}
\usepackage{twoopt} %for the tikz graohics which use two-option commands sometimes
\usepackage{ifthen}
\usepackage{upgreek}
\usepackage[utf8]{inputenc}
\usepackage{nicefrac} 
\usepackage{booktabs}
\usepackage{graphicx,float}
\usepackage{mathrsfs, amsfonts,bm}
\usepackage{epstopdf}
\usepackage{xcolor}

\usepackage[english]{babel}
\setcitestyle{square}

\newcommand{\qc}{q_\mathrm{c}}

\newcommand{\VC}{V_{\!\scriptscriptstyle\mathrm C}}

\newcommand{\n}{n}

\newcommand{\chiS}{\chi^0}

\newcommand{\chiSs}{\chiS_\sigma}

\newcommand{\chiSX}{\chi^0_{\scriptscriptstyle\mu=0}}
\newcommand{\chiSXbar}{\bar\chi^0_{\scriptscriptstyle\mu=0}}
\newcommand{\chiSmu}{\chi^0_{\scriptscriptstyle\mu\ne0}}
\newcommand{\chiSmubar}{\bar\chi^0_{\scriptscriptstyle\mu\ne0}}

\newcommand{\chinn}{\chi_{nn}}
\newcommand{\chiss}{\chi_{ss}}
\newcommand{\chins}{\chi_{ns}}

\newcommand{\qRd}{\bar{q}}

\newcommand{\vF}{v_{\scriptscriptstyle\mathrm{F}}}

\newcommand{\wRd}{\bar{\omega}}

\newcommand{\fermidist}{\mathrm{f}}

\newcommand{\grphDeg}{g_{\scriptscriptstyle\mathrm S}\,g_{\scriptscriptstyle\mathrm V}}
\newcommand{\kmu}{k_\mu}
\newcommand{\mus}{\mu_{\sigma}}
\newcommand{\kmus}{k_{\mu,\sigma}}
\newcommand{\must}{\mu_{\sigma\tau}}

\newcommand{\kms}{k^{\scriptscriptstyle(\mu)}_{\sigma}}
\newcommand{\muup}{\mu_{\uparrow}}
\newcommand{\mudo}{\mu_{\downarrow}}

\newcommand{\kcuoff}{k_\Lambda}

\definecolor{mydarkgreen}{RGB}{0,104,55}
\definecolor{mygreen}{RGB}{94,156,64}
\definecolor{mylightgreen}{RGB}{217,240,163}

\definecolor{mydarkblue}{RGB}{4,90,141}
\definecolor{myblue}{RGB}{10,54,126}
\definecolor{mylightblue}{RGB}{241,238,246}

\newcommand{\imgGrapheneZPlusRegions}[1][1.5]{
\begin{tikzpicture} [scale=#1]

\fill [mylightgreen, opacity=0.5]
      (2, 0) -- (3, 0) -- (3, 1) -- cycle;
\fill [mydarkgreen, opacity=0.5]
      (0,0) -- (2, 0) -- (3, 1) -- (3,3) -- (0,3) -- cycle;

\draw [thick, line width=0.2mm, dashed] (0, 0) -- (3, 3);
\draw [thick, line width=0.2mm, dashed] (0, 2) -- (2, 0);
\draw [thick, line width=0.2mm, dashed] (0, 2) -- (1, 3);
\draw [thick, line width=0.2mm] (2, 0) -- (3, 1);

\draw [thick, line width=0.3mm, -> ] (-0.1, 0) -- (3.2, 0) node [right] {$\bar{q}$};
\draw [thick, line width=0.3mm, -> ] (0, -0.1) -- (0, 3.2) node [above] {$\bar{\omega}$};
\foreach \i in {1,...,3} {
\draw [thick, line width=0.1mm] (\i, 0.05) -- (\i, -0.05) node [below] {$\scriptstyle \i$};
\draw [thick, line width=0.1mm, color=myblue] (0.05, \i) -- (-0.05, \i) node [left] {$\scriptstyle \i$};
}
\draw (-0.15, -0.15) node  {$\scriptstyle0$};

\draw (2.7,0.3) node  {$\scriptstyle z_+<1$};
\draw (1.9,1.0) node  {$\scriptstyle z_+>1$};
\end{tikzpicture}
}

\newcommand{\imgGrapheneZMinusRegions}[1][1.5]{
\begin{tikzpicture} [scale=#1]

\fill [mydarkgreen, opacity=0.5]
      (0,0) -- (2, 0) -- (0, 2) -- cycle;
\fill [mylightgreen, opacity=0.5]
      (0,2) -- 	(2, 0) -- (3, 0) -- (3, 2) -- cycle;
\fill [mylightblue, opacity=0.5]
      (0,2) -- 	(3, 2) -- (3, 3) -- (1, 3) -- cycle;      
\fill [mydarkblue, opacity=0.5]
      (0,2) -- 	(1, 3) -- (0,3) -- cycle;

\draw [thick, line width=0.2mm] (0, 2) -- (2.1, 2) node [right, baseline=center] {$\scriptscriptstyle z_-=0$};   \draw [thick, line width=0.2mm] (2.9, 2) -- (3, 2);
\draw [thick, line width=0.2mm, dashed] (0, 0) -- (3, 3);
\draw [thick, line width=0.2mm] (0, 2) -- (2, 0);
\draw [thick, line width=0.2mm] (0, 2) -- (1, 3);
\draw [thick, line width=0.2mm, dashed] (2, 0) -- (3, 1);

\draw [thick, line width=0.3mm, -> ] (-0.1, 0) -- (3.2, 0) node [right] {$\bar{q}$};
\draw [thick, line width=0.3mm, -> ] (0, -0.1) -- (0, 3.2) node [above] {$\bar{\omega}$};
\foreach \i in {1,...,3} {
\draw [thick, line width=0.1mm] (\i, 0.05) -- (\i, -0.05) node [below] {$\scriptstyle \i$};
\draw [thick, line width=0.1mm, color=myblue] (0.05, \i) -- (-0.05, \i) node [left] {$\scriptstyle \i$};
}
\draw (-0.15, -0.15) node  {$\scriptstyle0$};

\draw (1.9,1.0) node  {$\scriptstyle 0>z_-<1$};
\draw (0.8,0.3) node  {$\scriptstyle z_->1$};
\draw (1.2,2.3) node  {$\scriptstyle -1>z_-<0$};
\draw (0.38,2.8) node  {$\scriptstyle z_-<-1$};

\end{tikzpicture}
}
\definecolor{myGapheneConeColor}{RGB}{94,156,64}
\definecolor{myGapheneConeColorTwo}{RGB}{10,54,126}
\definecolor{myGapheneArrowColor}{RGB}{43,140,190}

\newcommand{\IcGrcMaxY}{1.3}
\newcommand{\IcGrcMinY}{-2}

\newcommand*\circled[2][0.0]{\tikz[baseline=#1]{
            \node[shape=circle,draw,inner sep=2pt] (char) {#2};}}

\newcommandtwoopt{\imgGrapheneCone}[3][1][0]{
\begin{tikzpicture} [scale=#1]
\draw [fill=white, color=white] (-2,-2.5) rectangle (1.5,2.5);

\draw [thick, line width=0.2mm, fill=myGapheneConeColor, opacity=0.4] (-1, -2) -- (1, -2) -- (0,0) -- cycle;
\draw [thick, line width=0.2mm, fill=myGapheneConeColor, opacity=0.4] ({-0.5*#3}, {#3}) -- ({0.5*#3}, {#3}) -- (0,0) -- cycle;
\draw [thick, line width=0.2mm] (-1, -2) -- (+1, +2);
\draw [thick, line width=0.2mm] (1, -2) -- (-1, +2);
\draw [thick, line width=0.3mm, -> ] (-1, 0) -- (1, 0) node [right] {$q$};
\draw [thick, line width=0.3mm, -> ] (0, -2) -- (0, 2) node [above] {$\epsilon(k)$};
\ifthenelse{\equal{#3}{0.0}}{}{
\draw  ({0.5*#3}, {#3}) node [right] {$\mu$};
}

\ifthenelse{\equal{#2}{1}}{
\draw[thick, line width=0.7mm, ->,>=stealth, color=myGapheneArrowColor](-0.75,-1.5) to [out=160,in=200] (-0.75, 1.5);

\ifthenelse{\equal{#3}{0.0}}{}{
\draw[thick, line width=0.7mm, ->,>=stealth, color=myGapheneArrowColor] ({0.25*#3}, {0.5*#3}) to [out=-70,in=-30] ({0.75*#3}, { 1.5*#3});
}

}{}

\end{tikzpicture}
}

\newcommand{\imgGrapheneConeSpin}[3][1.0]{
\begin{tikzpicture} [scale=#1]

\draw [fill=white, color=white] (-2,-2.5) rectangle (1.5,2.5);

\draw [thick, line width=0.2mm, fill=myGapheneConeColor, opacity=0.4] (-1, -2) -- (0, -2) -- (0,0) -- cycle;
\draw [thick, line width=0.2mm, fill=myGapheneConeColor, opacity=0.4] ({-0.5*#2}, {#2}) -- (0, {#2}) -- (0,0) -- cycle;
\draw [thick, line width=0.2mm, fill=myGapheneConeColorTwo, opacity=0.4] (1, -2) -- (0, -2) -- (0,0) -- cycle;
\draw [thick, line width=0.2mm, fill=myGapheneConeColorTwo, opacity=0.4] (0, {#3}) -- ({0.5*#3}, {#3}) -- (0,0) -- cycle;

\draw [thick, line width=0.2mm] (-1, -2) -- (+1, +2);
\draw [thick, line width=0.2mm] (1, -2) -- (-1, +2);
\draw [thick, line width=0.3mm, -> ] (-1, 0) -- (1, 0) node [right] {$q$};
\draw [thick, line width=0.3mm, -> ] (0, -2) -- (0, 2) node [above] {$\epsilon(k)$};
\ifthenelse{\equal{#3}{0.0}}{}{
\draw  ({0.5*#3}, {#3}) node [right] {$\mu_\uparrow$};}
\ifthenelse{\equal{#2}{0.0}}{}{
\draw  ({-0.5*#2}, {#2}) node [left] {$\mu_\downarrow$};}
\draw  (-0.3, -1.4) node [align=center, color=black] {$\downarrow$};
\draw  (0.3, -1.4) node [align=center, color=black] {$\uparrow$};

\end{tikzpicture}
}

\newcommandtwoopt{\imgGrapheneConeShift}[3][1][0]{
\begin{tikzpicture} [scale=#1]
\draw [fill=white, color=white] (-2,-2.5) rectangle (1.5,2.5);

\draw [thick, line width=0.2mm, fill=myGapheneConeColor, opacity=0.4] (0, \IcGrcMinY) -- ({(\IcGrcMinY+#3)/2}, {2*(\IcGrcMinY+#3)/2-#3}) -- (0, -#3) -- ({-abs(#3)/2}, 0) --  (0, 0) -- cycle;
\draw [thick, line width=0.2mm, fill=myGapheneConeColor, opacity=0.4] (0, \IcGrcMinY) -- ({-(\IcGrcMinY+#3)/2}, {2*(\IcGrcMinY+#3)/2-#3}) -- (0, -#3) -- ({abs(#3)/2}, 0) --  (0, 0) -- cycle; 

\draw [thick, line width=0.2mm] ({(\IcGrcMinY+#3)/2}, {2*(\IcGrcMinY+#3)/2-#3}) -- ( {(\IcGrcMaxY+#3)/2}, {2*(\IcGrcMaxY+#3)/2-#3});
\draw [thick, line width=0.2mm] ({-(\IcGrcMaxY+#3)/2}, {2*(\IcGrcMaxY+#3)/2-#3}) -- ( {-(\IcGrcMinY+#3)/2}, {2*(\IcGrcMinY+#3)/2-#3});
\draw [thick, line width=0.3mm, -> ] (-1, 0) -- (1, 0) node [right] {$q$};
\draw [thick, line width=0.3mm, -> ] (0, \IcGrcMinY) -- (0, \IcGrcMaxY*1.2) node [above] {$\epsilon(k)$};
\draw [dashed, line width=0.3mm] (0, -#3) -- (-0.5, -#3);
\draw [dashed, line width=0.3mm] (0, -#3) -- (0.5, -#3);
\ifthenelse{\equal{#3}{0.0}}{}{
\draw  ({0.5}, {-#3}) node [right] {$\mu$};
}

\ifthenelse{\equal{#2}{1}}{
\draw[thick, line width=0.7mm, ->,>=stealth, color=myGapheneArrowColor](-0.75,-1.5) to [out=160,in=200] (-0.75, 1.5);

\ifthenelse{\equal{#3}{0.0}}{}{
\draw[thick, line width=0.7mm, ->,>=stealth, color=myGapheneArrowColor] ({0.25*#3}, {0.5*#3}) to [out=-70,in=-30] ({0.75*#3}, { 1.5*#3});
}

}{}
\end{tikzpicture}
}

\newcommand{\imgGrapheneConeShiftSpin}[3][1.0]{
\begin{tikzpicture} [scale=#1]

\draw [fill=white, color=white,opacity=0.0] (-2,\IcGrcMinY*1.05) rectangle (1.5,{(-\IcGrcMinY+\IcGrcMaxY)*0.7});

\draw [thick, line width=0.2mm, fill=myGapheneConeColor, opacity=0.4] (0, \IcGrcMinY) -- ({(\IcGrcMinY+#2)/2}, {2*(\IcGrcMinY+#2)/2-#2}) -- (0, -#2) -- ({-abs(#2)/2}, 0) --  (0, 0) -- cycle;
\draw [thick, line width=0.2mm, fill=myGapheneConeColorTwo, opacity=0.4] (0, \IcGrcMinY) -- ({-(\IcGrcMinY+#3)/2}, {2*(\IcGrcMinY+#3)/2-#3}) -- (0, -#3) -- ({abs(#3)/2.0}, 0) --  (0, 0) -- cycle; 
\draw [thick, line width=0.2mm] ({(\IcGrcMinY+#2)/2}, {2*(\IcGrcMinY+#2)/2-#2}) -- (0,-#2) --  ( {-(\IcGrcMaxY+#2)/2}, {2*(\IcGrcMaxY+#2)/2-#2}); %left
\draw [thick, line width=0.2mm] ( {-(\IcGrcMinY+#3)/2}, {2*(\IcGrcMinY+#3)/2-#3}) -- (0,-#3) -- ( {(\IcGrcMaxY+#3)/2}, {2*(\IcGrcMaxY+#3)/2-#3}); %right
\draw [thick, line width=0.3mm, -> ] (-1, 0) -- (1, 0) node [right] {$q$};
\draw [thick, line width=0.3mm, -> ]  (0, \IcGrcMinY) -- (0, \IcGrcMaxY*1.2)  node [above] {$\epsilon(k)$};
\ifthenelse{\equal{#3}{0.0}}{}{
\draw  ({0.5}, {-#3}) node [right] {$\mu_\uparrow$};
\draw [dashed, line width=0.3mm] (0, -#3) -- (0.5, -#3);}
\ifthenelse{\equal{#2}{0.0}}{}{
\draw  ({-0.5}, {-#2}) node [left] {$\mu_\downarrow$};
\draw [dashed, line width=0.3mm] (0, -#2) -- (-0.5, -#2);}
\draw  (-0.3, \IcGrcMinY*0.9) node [align=center, color=black] {$\scriptstyle \downarrow$};
\draw  (0.3, \IcGrcMinY*0.9) node [align=center, color=black] {$\scriptstyle \uparrow$};

\end{tikzpicture}
}

\begin{document}

\title[Energy shift of collective modes in spin-imbalanced graphene on SiO\(_2\) from spin-sensitive linear response theory V.2]
{Energy shift of collective modes in spin-imbalanced graphene on SiO\(_2\) from spin-sensitive linear response theory}
% Force line breaks with \\
%\thanks{Footnote to title of article.}
\author{Dominik Kreil}
 \email{dominik.kreil@jku.at}
\author{Michaela Haslhofer}%
\author{Helga M. B\"ohm}
\affiliation{%
Institute for Theoretical Physics, Johannes Kepler University, Altenbergerstra\ss e 69, 4040 Linz, Austria%\\This line break forced% with \\
}%

\date{\today}

\begin{abstract}
%\cdjk{ca. 200 words}
The growing precision of optical and scattering experiments necessitates a better understanding of the influence of damping onto the collective mode of sheet electrons. As spin-polarized systems are of particular interest for spintronic applications, we here report spin-sensitive linear response functions of graphene, which give access to charge- and spin-density related excitations. We further calculate the reflectivity of graphene on an SiO\(_2\) surface, a setup used in s-wave scanning near field microscopy. Increasing the partial spin-polarization of the graphene charge carriers leads to a significant broadening  and shift of the plasmon mode, due to single-particle interband transitions of the minority spin carriers. We also predict an antiresonance in the longitudinal magnetic response function, similar to that of
semiconductor heterostructures.

\end{abstract}

%
%Valid PACS numbers may be entered using the \verb+\pacs{#1}+ command.
%\pacs{68.65.Pq} %Graphene films
%\pacs{05.30.Fk} Fermion systems and electron gas
%\pacs{71.45.Gm} Exchange, correlation, dielectric and magnetic response functions, plasmons
%\pacs{71.45.−d} Collective effects
%\pacs{71.10.−w} Theories and models of many-electron systems
%\pacs{71.10.Ca} Electron gas, Fermi gas
\maketitle

%%%%%%%%%%%%%%%%%%%%%%%%%%%%%%%%%%%%%%%%%%%%%%%%%%%%%%%%%%%%%%%%%%%%%%%%%%%%%%%%%%%%%%%%%%%
%%% ----------------------------------------------------------------------------------- %%%
%%%%%%%%%%%%%%%%%%%%%%%%%%%%%%%%%%%%%%%%%%%%%%%%%%%%%%%%%%%%%%%%%%%%%%%%%%%%%%%%%%%%%%%%%%%
%\begin{quotation}
%The ``lead paragraph'' is encapsulated with the \LaTeX\ 
%%\verb+quotation+ environment and is formatted as a single paragraph before the first section heading. 
%(%The \verb+quotation+ environment reverts to its usual meaning after the first sectioning command.) 
%Note that numbered references are allowed in the lead paragraph.
%
%The lead paragraph will only be found in an article being prepared for the journal \textit{Chaos}.
%\end{quotation}

\section{Introduction
\label{sec: Introduction}}

Although the existence of purely two-dimensional (2D) materials is prohibited by long-range thermal fluctuations~\cite{mermin1968crystalline}, Geim and Novoselov~\cite{novoselov2004electric} 2004  produced mono-atomic thin graphite layers, now famous as `graphene'. (Anharmonically coupled streching / bending modes~\cite{david2004statistical,le1992self, morozov2006strong} prevent the instability; the resulting rippling is avoided by placing the sheet on a flat support material).
Graphene's honeycomb structure implies many captivating properties, e.g.\ despite its thinness a mechanical strength 200 times that of steel. The large carrier mobility results in excellent thermal and electric conductivi ty, dynamically tunable by chemical doping or an applied gate voltage\cite{VakE11:transformation_optics}.
Doubtless, graphene is a promising candidate for high-speed and optoelectronic devices~\cite{yao2018broadband, garcia2014graphene}.

In addition, graphene holds a most intriguing potential for spintronic applications: The spin-orbit-coupling (SOC) allows to develop appliances where, due to their interplay, spin and charge currents can be manipulated simultaneously, offering the perspective of novel logic and memory devices.
Clearly, this requires a thorough theoretical understanding of graphene's spin-resolved properties. Of particular interest is the collective behavior of the charge carriers: Effective, spin-dependent interactions and correlations between the charge carriers (electrons or holes) have manifest fingerprints in the excitation spectrum, accessible experimentally. Specifically, light \textit{scattering} from surfaces using \textit{scanning near-field optical microscopy} (\textit{s-SNOM}) has provided accurate data on graphene, pioneered by Fei et al.~\cite{fei2011infrared} in the mid-infrared, and later extended to the teraherz range~\cite{yan2013damping,low2014graphene}. 

Graphene's valence- and conduction-band energies touch at the 6 corner points of the Brillouin zone (BZ). Half of them are equivalent (as two atoms are in the unit cell), and referred to as \(K\) and \(K'\) points. In their vicinity electrons and holes behave as massless Dirac Fermions~\cite{neto2009electronic} with a \textit{linear} energy dispersion, contrary to the quadratic one of conventional 2D electron liquids (2DELs) in semiconductor layers.
The single-particle excitations then form continua (`particle-hole bands, PHBs) with linear boundaries; they offer prominent decay channels
(`Landau damping'~\cite{landau1946vibrations}) for the plasmon.
The vanishing gap brings about another crucial difference to the standard interface 2DELs: \textit{interband} transitions lead to damping at much shorter wavelengths than the intraband PHB. 
Consequently, graphene's plasmon is much stronger influenced by a spin polarization of the system, because the interband PHB edge is drastically decreased with increasing spin imbalance.

Experimentally, 2DELs with a different amount of \(\uparrow\) and \(\downarrow\) spins have been realized (various methods being reviewed in~\cite{wolf2001spintronics}). For such systems a long-lived `spin-plasmon' (or longitudinal magnon) was predicted~\cite{APVF14:long-lived}. Placing such a spin-imbalanced 2DEL between coupled (spin-torque) nano-magnets,  would enable to tune their coupling via controlling the spin populations. 
With the prominent electron layers realized in GaAs-GaAlAs heterostructures this intriguing idea does not work, as correlations lower the spin-plasmon peak \cite{kreil2015excitations}. 

In graphene, where the touching Dirac cones imply a richer excitation spectrum even in the simple linear-dispersion model, prevent a straight-forward generalization of these predictions. Therefore, it is highly interesting to study the effect from scratch.
To the best of our knowledge, the dielectric response of partially spin-polarized graphene has not yet been investigated. In this work, we derive the partial Lindhard functions from spin-sensitive linear response theory.
For ease of reading, the spin-density (\(\propto\) the magnetization's \(z-\)component) in this context is simply referred to as `spin', i.e.\ `spin--spin response' stands for `spin-density--spin-density response'. 
We performed the first Random Phase Approximation (RPA) calculations for single-layer graphene with a spin-imbalance, and present
results for the density--density (or charge--charge) response, the spin--spin response, as well as for the density--spin response. The latter describes magnetic excitations caused by electric perturbations and vice-versa, resulting not from SOC but Coulomb interaction and Pauli exclusion. 

In addition, we apply our dielectric function to determine the reflectivity of graphene on a SiO\(_2\) substrate as studied in s-SNOM experiments~\cite{fei2011infrared}.
When, at sufficiently high doping, the plasmon energy reaches that of optical phonons in the substrate, the coupling between the modes causes the dispersions to `repel' each other. We investigate how this is affected by a spin-imbalance.
 
This work is organized as follows:
We first address in Sec.\,\ref{sec: Collective Modes} the fundamental ambiguity of defining a collec\-tive mode's precise location, if damping is significant. In Sec.\,\ref{sec: Spin-Sensitive Linear Response Theory}, after briefly reviewing the energy bands and presenting the spin-dependent formalism (\ref{ssec: Energy Dispersion near the Dirac Points}), we derive the partial Lindhard functions for spin \(\sigma\) fermions (\ref{ssec: Partial Lindhard Functions of Graphene}) and the resulting RPA response functions (\ref{ssec: Spin-Sensitive Random Phase Approximation}).
The reflectivity of spin-imbalanced graphene on SiO\(_2\) is studied in Sec.\,\ref{sec: SNOM}, followed by a critical discussion of our results in Sec.\,\ref{sec: Conclusions}.
All calculations are done for zero temperature \(T=0\), the majority spins are denoted without loss of generality as \(\uparrow\). 

%%%%%%%%%%%%%%%%%%%%%%%%%%%%%%%%%%%
\section{Collective Modes
 \label{sec: Collective Modes}}

Collective excitations of many particles are characterized by their in-phase movement. Charge carriers in solids have additional degrees of freedom, e.g.\ the spin, and possibly a `pseudospin' due to different valleys (band structure minima with the same energy at different points of the BZ as in graphene's \(K,\,K'\) points).
In such multi-component systems, in addition to the overall collective mode of the density, the various sub-species may oscillate with opposite phases. 
The full density mode is the plasmon (with an in-phase oscillation of all spins). 
The longitudinal mode where \(\uparrow\!-\)spins collectively move against \(\downarrow\!-\)spins is referred to as the `spin-plasmon'\cite{APVF14:long-lived}.

Long-lived collective excitations are mathematically found from poles in response functions, equivalent to peaks in the scattering cross section. However, if damping and drag forces are present, their determination becomes ambiguous~\cite{HDKB17_Phenomenplbroadening}. We exemplify this for the Drude model for classical charge carriers~\cite{pines2018elementary}: \(\epsilon_\mathrm{D}(\omega)\!=\!1-\omega_\mathrm{pl}^2/\omega(\omega+i\bar\eta) \), with the classical plasmon frequency \(\omega_\mathrm{pl}\) and damping parameter \(\bar\eta\!\equiv \eta\,\omega_\mathrm{pl}\).
Measurements~\cite{lee2011optical} of graphene's optical transmission and reflection coefficient are well described by this model with \(\eta\!=0.007\). 
From this spectroscopic perspective, the plasmon is best defined~\cite{bonitz1998quantum} as the \textit{complex} zero of the complex dielectric function \(\epsilon(\omega_1\!+\!i\omega_2)\).
In the Drude model damping shifts the observed mode towards lower energies \(\,\omega_1/\omega_\mathrm{pl}= (1-\eta^2/4)^{1/2}\), with \(\omega_2\!=\eta/2\).

By contrast, scattering experiments (e.g.\ electron energy loss spectroscopy (EELS)~\cite{hirjibehedin02evidence, perez2009spin}) probe the loss function \(\mathrm{Im}\,\epsilon^{-1}(\omega)\), proportional to 
the scattering cross-section. %\(\partial^2 \sigma/\partial \hbar\omega \partial \Omega\). 
Correspondingly, the plasmon is defined as a peak in the latter. For small \(\eta\) the two definitions agree nicely, and, in addition, \(\omega_1\) (the real part of the complex root of \(\epsilon\)) is also very close to the zero of \(\mathrm{Re}\,\epsilon\) (a plasmon definition found in many textbooks).

\begin{figure}[H]
\pgfimage[width=0.5\textwidth]{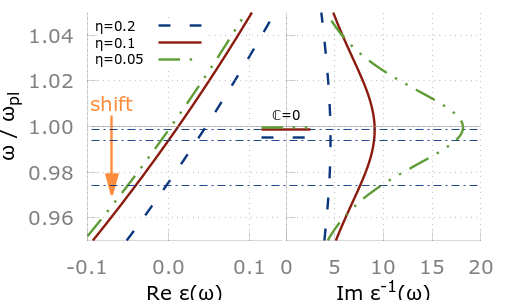}
\caption{Real Drude dielectric function (left). The zero of \(\mathrm{Re}\,\epsilon(\omega)\) decreases rapidly with damping rate \(\eta\!=\bar\eta/\omega_{\mathrm{pl}}\).
%three different damping rates \(\bar\eta/\omega_{\mathrm{pl}}=0.05\),  0.1 and 0.2. 
The plasmon observed in EELS is at the maximum of \(\mathrm{Im}\,\epsilon^{-1}(\omega)\) (right). The complex root of \(\epsilon(\omega)\) 
(short line in the middle, marked `\(\mathbb{C}\!=\!0\)') is affected much less. \label{fig:drude_loss_function}}
\end{figure}

With increasing wave-vector \(q\!\ne\!0\) the graphene plasmon enters the inter-band single-particle continuum and gets highly Landau-damped.
For large \(\eta\), using the appropriate definition for calculating the collective modes becomes crucial~\cite{HDKB17_Phenomenplbroadening}.
In Fig.\ref{fig:drude_loss_function}, we show the real and imaginary part of the Drude loss function. While the maximum of \(\mathrm{Im}\,\epsilon^{-1}_\mathrm{D}(\omega)\) is hardly effected by a damping of \(\eta\!\lesssim\!1\), the zero of \(\mathrm{Re}\,\epsilon_\mathrm{D}(\omega)\) is significantly lowered. 
%\chmb{shown/or not}

This sensitivity on \(\eta\) shows the importance of using the appropriate definition for the collective modes in spin-imbalanced graphene, where interband damping is formidable.  
We therefore obtain the plasmon from the maxima in the imaginary part of the response functions (condition of maximal dissipation \cite{fei2011infrared}).
Before presenting our results of these loss functions and that of the influence of the substrate, we briefly review the linear response theory for a homogeneous and isotropic 2DEL with linear energy dispersion, and derive the spin-resolved RPA response functions.

%%%%%%%%%%%%%%%%%%%%%%%%%%%%%%%%%%%
\section{Spin-Sensitive Linear Response Theory
 \label{sec: Spin-Sensitive Linear Response Theory}}

%%%%%%%%%%%%
\subsection{Energy Dispersion near the Dirac Points
 \label{ssec: Energy Dispersion near the Dirac Points}}

The Pauli matrices \(\underline{\underline{\mbox{\boldmath$\sigma$}}}\) describe the particle's behavior in an applied magnetic field  \({\bf B}\), with \(\underline{\underline{\sigma_z}}\) denoting their orientation with respect to a given direction. 
Similarly, two more Pauli vectors, termed `pseudospin', \(\underline{\underline{\mbox{\boldmath$\tau$}}}^{\scriptscriptstyle K}\)
and \(\underline{\underline{\mbox{\boldmath$\tau$}}}^{\scriptscriptstyle K'}\) describe the states at \(K\) and \(K'\).
Single-particle energies \(\varepsilon^\ell_{\sigma,\tau,{\bf k}}\) are thus characterized by their band-index \(\ell\), spin and pseudospin \(\sigma,\tau\), and a 2D wave vector \({\bf k}\). The latter is measured relative from \(K\) and \(K'\), respectively.

A thorough first-principles study of graphene's band structure in the presence of SOC was presented in Refs.~\cite{AEHM10:spin-split,FGOM13:impact}. Aiming at spintronics applications, a large external (or `Rashba') SOC induced by an external electric field or magnetic adatoms is desirable. Although giving rise to a band splitting near the Dirac points, this gap is rather small~\cite{AEHM10:spin-split}. Therefore, using a linear energy dispersion in vicinity of \(K,K'\) is a good starting point.

Denoting with \(\mu\) the chemical potential and 
measuring all energies with respect to the Fermi energy \(\mu(T\!=\!0)\) of the undoped graphene (i.e.\ at the meeting point of the upper and lower Dirac cone)
the single-particle Hamiltonian of a charge carrier with Land\'e factor 2 can then near \(K\) be expressed as~\cite{neto2009electronic}
\begin{subequations}\begin{equation}
   \underline{\underline{\hat{\bar h}}}_{\sigma,\scriptscriptstyle K} \equiv\> 
         \underline{\underline{\hat{h}}}_{\scriptscriptstyle K} \!+ 
         \underline{\underline{\sigma_z}}\,\mu_{\scriptscriptstyle\mathrm B}B_z - 
         \mu\, \;,
%                 \mu\,\underline{\underline{1}} \;,
 \label{equ:DiracHamiltonian}
\end{equation}
where \(\mu_{\scriptscriptstyle\mathrm B}\) is Bohr's magneton and unity matrices are not spelled out explicitly. The pure graphene part, in compact and in matrix form, reads  
\begin{equation}
   \underline{\underline{\hat{h}}}_{\scriptscriptstyle K} =\>
   \hbar\vF\,\hat{\mathbf{p}}\!\cdot\! \underline{\underline{\mbox{\boldmath$\tau$}}}^{\scriptscriptstyle K} 
   \>=\> \hbar\vF\,\mbox{\small\(
 \begin{pmatrix} 0 &\!\! \frac1{i}\frac{\partial}{\partial x} \!+\! \frac{\partial}{\partial y}\\
    \frac1{i}\frac{\partial}{\partial x} \!-\! \frac{\partial}{\partial y} \!\! &0            
 \end{pmatrix}\)} \;.
 \label{equ:DiracH matrix}
\end{equation}
where \(\vF\!\approx 10^6\)m/s \cite{neto2009electronic} is a material constant and \(\hat{\mathbf{p}}=\hbar\nabla/i\) the 2D momentum operator.
\end{subequations}
A similar operator \(\hat{h}_{\sigma,\scriptscriptstyle K'}\) holds near \(K'\) (see Appendix \ref{app: H in Kprime}).

\begin{figure}[h]
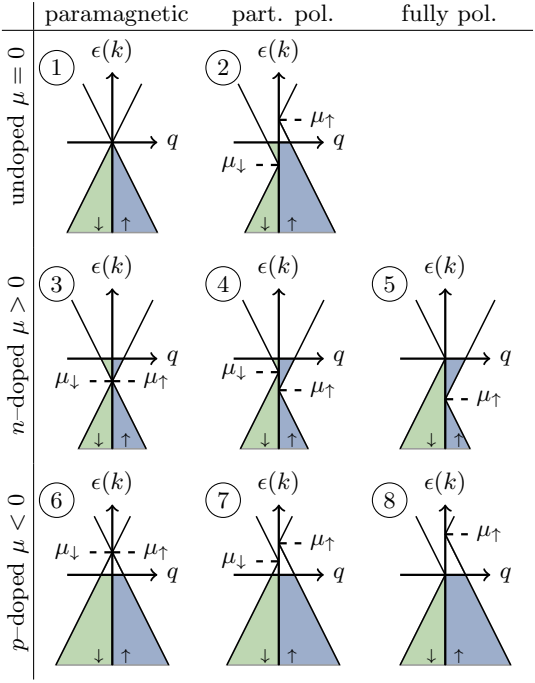

\centering

\begin{tabular}{r|ccc}
& paramagnetic & part. pol. & fully pol. \\ \hline
\rotatebox{90}{\hspace{1em} undoped \(\mu=0\)} &\circled[-7em]{1}\hspace{-2.5em}\imgGrapheneConeShiftSpin[0.6]{0.0}{0.0}  & \circled[-7em]{2}\hspace{-2.5em}\imgGrapheneConeShiftSpin[0.6]{0.5}{-0.5} \\
\rotatebox{90}{\hspace{0.5em} \(n\)--doped \(\mu>0\)} & \circled[-7em]{3}\hspace{-2.5em}\imgGrapheneConeShiftSpin[0.6]{0.5}{0.5} & \circled[-7em]{4}\hspace{-2.5em}\imgGrapheneConeShiftSpin[0.6]{0.3}{0.7} & \circled[-7em]{5}\hspace{-2.5em}\imgGrapheneConeShiftSpin[0.6]{0.0}{0.9} \\
\rotatebox{90}{\hspace{0.5em} \(p\)--doped \(\mu<0\)} & \circled[-7em]{6}\hspace{-2.5em}\imgGrapheneConeShiftSpin[0.6]{-0.5}{-0.5} & \circled[-7em]{7}\hspace{-2.5em}\imgGrapheneConeShiftSpin[0.6]{-0.3}{-0.7} & \circled[-7em]{8}\hspace{-2.5em}\imgGrapheneConeShiftSpin[0.6]{0.0}{-0.9}
\end{tabular}

%\imgGrapheneCone[0.5]{0.0}\imgGrapheneCone[0.5]{1.0}\imgGrapheneConeSpin[0.5]{0.5}{1.5}\imgGrapheneConeSpin[0.5]{0.0}{1.9}
\caption{Energy dispersions for all 8 possible cases: undoped and doped graphene with paramagnetic, partially polarized and fully polarized setting. The left (green) cones depict the minority (or spin \(\downarrow\)) charge carriers, and the right (blue) cones the majority (or spin \(\uparrow\)) ones. Consistent with Eq.\eqref{equ:DiracEnergy}, \(\mus \mbox{\tiny$\begin{matrix} \\[-1.9em] < \\[-0.6em]\scriptscriptstyle (>)\end{matrix}$}0\) shifts the crossing point to lower (higher) energies.    \label{fig:free_cones}}
\end{figure}

Both, \(\hat{h}_{\scriptscriptstyle K}\) and \(\hat{h}_{\scriptscriptstyle K'}\) yield the same energy dispersion proportional \(k\!=\!|{\bf k}|\) resulting in the eigenvalues of Eq.\eqref{equ:DiracHamiltonian}) 
\begin{subequations}\begin{align}
  \epsilon^\ell_{\sigma\tau}(k) &=\> 
  \ell\,\hbar\vF\, k + \sigma\mu_{\scriptscriptstyle\mathrm B} B_z - \mu \,,
  \\
  &\equiv\> \ell\,\hbar\vF\, k - \mus \;.
\end{align}\label{equ:DiracEnergy}\end{subequations}
The only effect of the valleys being therefore to contribute a degeneracy factor \(g_{\scriptscriptstyle\mathrm V}\!=\!2\) in summations, we suppress this index in the following.
The dispersion Eq.\eqref{equ:DiracEnergy} suggests the definition of spin-dependent chemical potentials \(\mus\) as explained in Fig.\ref{fig:free_cones}.
These determine the maximal wave vectors \(\kms \!\equiv\! \mus/\hbar\vF\) for occupations with spin \(\sigma\) (the `Fermi wave vectors' of each spin component). 
Without a magnetic \(B-\)field, in undoped graphene all \(\,\mus\!= 0=\!\mu \) with no electrons in the conduction and no holes in the valence band. A system in \(B\!\ne\!0\) has at least one \(\mus\!\neq\! 0\). 

The density of charge carriers with spin \(\sigma\) in the conduction and valence  band determines the \(\must\)
via the \(T\!=\!0\) Fermi distribution function \(\fermidist(\varepsilon)\) and the 
energies of Eq.\eqref{equ:DiracEnergy}
\begin{equation} \label{equ: ns to mus}
 \begin{array}{lllll}
  n^\ell_\sigma &\!\!=\; \sum\limits_\tau \int\!\!\frac{d^2k}{(2\uppi)^2}\> 
           \fermidist\big(\varepsilon^\ell_{\sigma\tau}(\mathbf{k}) \big)
           \\
       &\!\!=\; \frac{g_{\scriptscriptstyle\mathrm V}}{2\uppi}
           \int\!kdk\> \theta\big(\mus\!-\!\ell\hbar\vF\,k\big)
           \end{array} \;.
\end{equation}
For the partial chemical potentials \(\mus\) there are eight different scenarios possible, corresponding to a system with the following properties:
\begin{enumerate}\itemsep=-0.2em
  \item undoped, paramagnetic: \(\muup =\!0\!= \mudo\), both spin-species have the same density \(n_{\uparrow}\!=\!n_{\downarrow}\)
  \item undoped, partially polarized: \(\muup = -\mudo\), \(n_{\uparrow}\!>\!n_{\downarrow}\)
    \vspace{0.5em}
  \item \(n-\)doped, paramagnetic: \(\muup \!=\! \mudo \!>\!0\), \(n_{\uparrow}\!=\!n_{\downarrow}\)
  \item \(n-\)doped, partially polarized: \(\muup \!>\! \mudo\!>\!0\), \(n_{\uparrow}\!>\!n_{\downarrow}\)
  \item \(n-\)doped, fully polarized: \(\muup \!>\!0\), \(\mudo \!=\! 0\), \(n_{\downarrow}\!=\! 0\)
     \vspace{0.5em}
  \item \(p-\)doped, paramagnetic: \(\muup \!=\! \mudo \!<\!0\), \(n_{\uparrow}\!=\!n_{\downarrow}\)
  \item \(p-\)doped, partially polarized: \(\muup \!<\! \mudo\!<\!0\), \(n_{\uparrow}\!>\!n_{\downarrow}\)
  \item \(p-\)doped, fully polarized: \(\muup \!>\!0\), \(\mudo \!=\! 0\), \(n_{\downarrow}\!=\! 0\).
\end{enumerate}
In Fig.\ref{fig:free_cones} these eight cases are depicted schematically. From diagrams 5 and 8 it is seen that in these cases interband excitations are possible with zero energy.

With hindsight to spintronic applications the spin-imbalanced doped cases are of major interest. Without SOC, all the magnetic properties discussed here, in \(p-\) and \(n-\)doped graphene behave in exactly the same manner. Without loss of generality we therefore assume the valence band to be full. 

The total density now equals that of the conduction band, and defines the cut-off wave vector \(\kmu\) as that of the paramagnetic system with this density (spin degeneracy factor \(g_{\scriptscriptstyle\mathrm S}\!=\!2\)). The maximally occupied \(\kms\) obey analogous relations,
\begin{subequations}\begin{eqnarray}
  n =  \sum_\sigma n^{^{\mathrm{C}}}_{\sigma} &\equiv&  \textstyle\frac{\grphDeg}{4\uppi}\,\displaystyle\kmu^2 
    = \textstyle\frac{\grphDeg}{4\uppi\,(\hbar\,\vF)^2}\displaystyle \mu^2 \;,
    \label{equ: kmu def}
  \\
  n_\sigma = n^{^{\mathrm{C}}}_{\sigma} &\equiv&  \textstyle\frac{g_{\scriptscriptstyle\mathrm V}}{4\uppi}\,\displaystyle\kmu^2
   = \textstyle\frac{g_{\scriptscriptstyle\mathrm V}}{4\uppi\,(\hbar\,\vF)^2}\displaystyle \mus^2 \;.
   \label{equ: kmus def}
\end{eqnarray}\end{subequations}

The polarization parameter \(\zeta\) quantifies the spin imbalance in partially spin polarized systems (\(\bar\sigma\!\equiv -\sigma\) denotes the opposite spin)
\begin{equation}
\zeta_\sigma \equiv  \left( \n_\sigma - \n_{\bar{\sigma}} \right)/n  \ .  \label{equ:def_zeta_sigma}
\end{equation}
Inverting this, one readily finds \(\n_{\sigma}\!=(1\!+\!\zeta)\n/2\) and the relation of the partial Fermi wave vectors \(\kmus\) and energies \(\mus\) with those of the paramagnetic system
\begin{equation}
 \begin{matrix} \kmus = \sqrt{1\!+\!\zeta_\sigma}\,\kmu \ \\ 
                 \mus =  \sqrt{1\!+\!\zeta_\sigma}\,\mu  \end{matrix} \;.
\end{equation}
We conclude this section with noting that in order to achieve a finite density, an empirical cut-off parameter  \(\kcuoff\) must be introduced for the valence band such that the number of states in the BZ is conserved~\cite{dharma2006coulomb}. In the area \(A_{\scriptscriptstyle\mathrm P}\!=3a^2/2\) (\(a\!=14.2\)nm~\cite{cooper2012experimental}) of the primitive cell, each carbon atom contributes one p-orbital state to the valence band (as well as an other to the conduction band). This corresponds to defining \(\kcuoff\) from the density 
\(\n^{\scriptscriptstyle\mathrm V}= 2/A_{\scriptscriptstyle\mathrm P} \equiv \frac{g_{\scriptscriptstyle\mathrm S}g_{\scriptscriptstyle\mathrm V}}{4\uppi}\kcuoff^2 \) as a constant of the system.

\newpage
\subsection{Partial Lindhard Functions of Graphene
 \label{ssec: Partial Lindhard Functions of Graphene}}

The free polarizability (or `Lindhard function') of graphene is given by~\cite{wunsch2006dynamical}
(where \(\fermidist^\ell_{\sigma}(\mathbf{k}) \equiv  \fermidist(\varepsilon^\ell_{\sigma}(\mathbf{k}))\,\) are the Fermi functions)
\begin{align}
 \nonumber\chiS_{\sigma}(q,\omega) &= 
 \frac{g_{\scriptscriptstyle\mathrm V}}{\Omega}\sum\limits_{\ell,\ell',\mathbf{k}} 
 \frac{\fermidist^\ell_{\sigma}(\mathbf{q})-\fermidist^{\ell'}_{\sigma}(\mathbf{k}+\mathbf{q})}
      {\hbar\omega -\epsilon^{\ell'}_{\sigma}(\mathbf{k}\!+\!\mathbf{q}) +\epsilon^\ell_{\sigma}(k) + i\,0^+}  \\ &\times\Big|\big\langle \phi^\ell_{\sigma}(\mathbf{k}) \big| \phi^{\ell'}_{\sigma}(\mathbf{k}\!+\!\mathbf{q}) \big\rangle \Big|^2
\end{align}
with the small imaginary part in the denominator ensuring causality, \(\Omega\) denoting the volume, and
\(\big|\phi^{\ell}_{\sigma}(\mathbf{k})\big\rangle\) the eigenstates of the Hamiltonian in Eq.\eqref{equ:DiracHamiltonian} for band index \(\ell\!\in \scriptstyle\lbrace{\mathrm{C,V}\rbrace}\). The full density-density response function is obtained by \(\chiS(q,\omega) \equiv \chi^0_{nn}(q,\omega)  = \sum\limits_{\sigma} \chiS_{\sigma}(q,\omega) \). 

In the following, we successively present the undoped and doped paramagnetic  \(\chiS\) ~\cite{wunsch2006dynamical} and then derive the partially spin-polarized Lindhard functions.
All quantities are given in reduced units, energies being measured in \(\mu\) and lengths in \(k_\mu\) of the paramagnetic 2D Dirac liquid, in particular
\(\qRd\!=q/\kmu\), \(\wRd\!= \hbar\omega/\mu\), and \(\bar\chi\!= \mu\chi/n\). 
The alert reader may notice that these appear to diverge in the undoped system where \(\mu\!\rightarrow\!0\). In this case, any arbitrary \(\tilde\mu\) and \(\tilde k_\mu\) with \(\tilde\mu\!=\!\hbar\,\vF\tilde k_\mu\) describes the same result. This reflects the fact that \(\chi^0(q,\omega)\propto n/\mu\) which both vanish, but at a finite ratio.
The chosen units allow meaningful comparisons of systems with different \(\zeta\) in analogy to conventional 2DELs with parabolic dispersion.

\paragraph{Undoped system (\(\mu = 0 \)):}

In the case of no doping, where all \(\mus\!=\!0\), the full free response reads~\cite{wunsch2006dynamical}
\begin{subequations}\begin{align}
\nonumber\chiSX(q, \omega) &= \frac{\grphDeg}{16}  \frac{q^2}{\hbar\vF\sqrt{q^2 - (\omega/\vF)^2}} \\
&= \frac{\grphDeg}{16}  
   \frac{\kmu^2}{\mu} \frac{ (q/\kmu)^2  }
        { \sqrt{(q/\kmu)^2 - (\omega/\mu)^2 } } \;,
\end{align}
or, in reduced units,
\begin{equation}
 \chiSXbar(\qRd, \wRd) = \frac{1}{32\uppi}\, \frac{\qRd^2}{\sqrt{\qRd^2 - \wRd^2}}
 \,\equiv\,  \frac{\n}{\mu}\,\chiSX(\qRd, \wRd) \;.
 \label{equ:graohene_undoped_response_reduced} 
\end{equation}\end{subequations}
The partial spin response is \(\chiS_{\sigma}\!=\!\tfrac{1}{g_{\scriptscriptstyle\mathrm S}}\chiS\) due to symmetry.
\medskip

\paragraph{Doped System (\(\mu\!\neq\!0 \)):}

\begin{figure}[h]
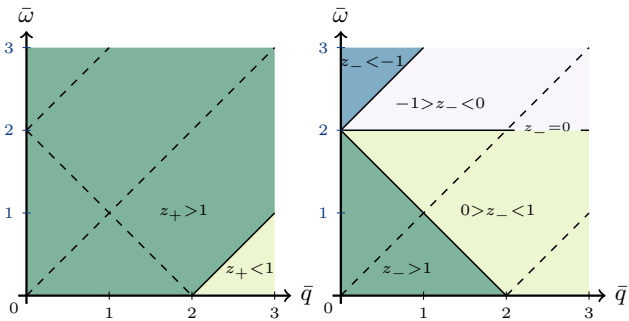

\centering
\imgGrapheneZPlusRegions[1.1]\hspace{-10pt}\imgGrapheneZMinusRegions[1.1]
\caption{Regions in the reduced \((\qRd, \wRd)\) plane  in which \(z_+\) (left) and \(z_-\) (right) are positive (green), negative (blue), and where they exceed \(\pm1\).  \label{fig:regions_for_complex_discussions} }
\end{figure}

A non-vanishing chemical potential \(\mus\!\neq\!0\) changes the form of the response function dramatically to
\begin{align}
\nonumber\chiSmu(q, \omega)&=\> \frac{\grphDeg}{16\uppi}  
 \frac{\kmu^2}{\mu} \!\left(\frac{ \qRd^2\, \mathcal{F}(\qRd,\wRd) }{\sqrt{\qRd^2 - \wRd^2}}-8  \right)   \;,
 \\
\chiSmubar(\qRd, \wRd) &= - \frac{1}{4\uppi^2} - \frac{1}{32\uppi^2} \frac{\qRd^2\, \mathcal{F}(\qRd,\wRd)}{\sqrt{\qRd^2 - \wRd^2}} \;,
 \label{equ:graohene_doped_response_reduced}
\end{align}
with 
\begin{equation}
\mathcal{F}(\qRd,\wRd) \>=\>
 \mathcal{G}^+\!\Big(\mbox{\small\(\displaystyle\frac{2+\wRd}{\qRd}\)}\Big)  - 
 \mathcal{G}^-\!\Big(\mbox{\small\(\displaystyle\frac{2-\wRd}{\qRd}\)}\Big)
\end{equation}
and \(\mathcal{G}^\pm(z) = z\sqrt{1-z^2}\pm i\,\mathrm{arccosh}(z)\). The function \(\mathcal{F}\) determines the structure of the response function in the (\(\qRd, \wRd\))-plane, the various arising regions characterized by \(z_\pm\!\equiv (2\pm\wRd)/\qRd\) are shown in Fig.\,\ref{fig:regions_for_complex_discussions} . 

In a spin-polarized system, the response functions of the constituents are rescaled with the individual Fermi-wave vectors \(\kmus\). Thus we get for the partial response functions 
\begin{align}
\nonumber\chiSs(q, \omega;\,\mu)  &= \left\lbrace \begin{matrix}
 \frac{\n}{\mu\,g_{\scriptscriptstyle\mathrm S}}\,\chiSXbar(\qRd, \wRd) 
    &\text{for }\> \mu_{\sigma}\!=\!0  & \text{(undoped)} 
 \vspace{0.15cm}\\  
  \,\frac{\n_{\sigma}} {\mu_{\sigma}}\,\chiSmubar(\qRd_{\sigma}, \wRd_{\sigma}) 
    &\text{for }\>  \mu_{\sigma}\!\neq\!0  & \text{(doped)} 
 \end{matrix} \right. \\
&\equiv^{\phantom{\Big|}\!} \bar{\chi}^0_{\sigma}(\qRd, \wRd)\,\n/\mu \ .
%&\equiv^{\phantom{\Big|}\!} \textstyle{\frac{\n} {\mu}}\,\bar{\chi}^0_{\sigma}(\qRd, \wRd) \ .
\end{align}
%= \frac{\n_{\sigma\tau}} {\mu_{\sigma\tau}}\,\bar{\chi}^0(\qRd_{\sigma\tau}, \wRd_{\sigma\tau})
%This leads for the reduced spin--valley polarized response to
%\begin{equation}
%\bar{\chi}^0_{\sigma\tau}(\qRd, \wRd)  =  \left\lbrace \begin{matrix} \frac{\sqrt{\left(1+\zs \right)\,\left(1+\zv \right)}}{4} \,\bar{\chi}^0(\frac{\qRd}{\sqrt{\left(1+\zs \right)\,\left(1+\zv \right)}}, \frac{\wRd}{\sqrt{\left(1+\zs \right)\,\left(1+\zv \right)}})  &  \text{for } \mu_{\sigma\tau} \neq 0  \\  \frac1{4}\,\bar{\chi}^0_\text{X}(\qRd, \wRd) &  \text{for } \mu_{\sigma\tau} = 0 \end{matrix} \right.
%\end{equation}

The Fourier transform of the Coulomb interaction expressed in reduced units is \( \VC(q) = \frac{\mu}{\n}\bar{\VC}(\qRd) = 2\uppi\hbar\vF\alpha/q\) with the effective graphene coupling constant \(\alpha\). For a free standing graphene sheet it is given by~\cite{wunsch2006dynamical} \(\alpha\!=\!\alpha_{\scriptscriptstyle\mathrm{gr}}\!=\!e^2/4\uppi\epsilon_0\hbar\vF\approx 2.2\); and by \(\alpha\!=\!\alpha_{\scriptscriptstyle\mathrm{gr}}/\kappa_{\scriptscriptstyle\mathrm S}(\omega)\) in a surrounding with a dielectric function \(\kappa_{\scriptscriptstyle\mathrm S}(\omega)\). 
\goodbreak

%%%%%%%%%%%%
\subsection{Spin-Sensitive Random Phase Approximation
 \label{ssec: Spin-Sensitive Random Phase Approximation}}

The observed excitations are determined by mutually dependent exchange and correlation effects.
In the RPA, particles react with the free response \(\chiSs\) (ensuring the Pauli principle and thus accounting for exchange) to the effective electric and magnetic field in the probe; this mean field reflects the correlations. Generalized RPA theories use refined effective interactions \(V_{\sigma\sigma'}(q)\), various spin-dependent potentials having been introduced for the 2DEL~\cite{Iwam91, giuliani2005quantum, kreil2015excitations,asgari2006static, moreno2003local}.
The generalized RPA response of multi-component systems takes a matrix from ~\cite{giuliani2005quantum}
\begin{equation}
\bm{\chi}  = (\mathbf{1} - \mathbf{V}\cdot\bm{\chi}^0 )^{-1}\cdot\bm{\chi}^0   \;. \label{equ:chi_matrix_exact}
\end{equation}
In the present case \(\mathbf{V}\!=\!\big(V_{\sigma\sigma'}\big)\), and \(\bm{\chi}^0\!=\!\big(\delta_{\sigma\sigma'}\chiSs\big)\). By generalizing these potentials further to dynamic ones, allows to capture double plasmon excitations \cite{BHKP10:dynamic, PaGR18:nonlocal} and intrinsic damping. Except for artificial graphene~\cite{GSPP09:engineering, PGLM13:artificial, WSDK18:observation}, correlations here~\footnote{The parameter \(r_{\scriptscriptstyle\rm S}\), termed \(\alpha\) in graphene, has the constant value of approx. 2.2.},~\cite{liu2008plasmon} are well described by the `bare' RPA, where all matrix elements are the Coulomb potential \(\VC(q)\).

We calculated the density--density, density--spin and spin--spin response function, which can be obtained from Eq.\eqref{equ:chi_matrix_exact}
\begin{subequations}\begin{eqnarray}
\chinn &= \sum\limits_{\sigma, \sigma'} \chi_{\sigma, \sigma'} 
       &=\> \chi_{\uparrow\uparrow} + \chi_{\downarrow\downarrow} + 2\chi_{\uparrow\downarrow} \;,  \\
\chiss &= \sum\limits_{\sigma, \sigma'} \,\sigma\sigma'\,\chi_{\sigma, \sigma'}
       &=\> \chi_{\uparrow\uparrow} + \chi_{\downarrow\downarrow} - 2\chi_{\uparrow\downarrow} \;,  \\     
\chins &= \sum\limits_{\sigma, \sigma'} \>\sigma\> \chi_{\sigma, \sigma'} 
        &=\>\chi_{\uparrow\uparrow} - \chi_{\downarrow\downarrow} \;.
\end{eqnarray}\end{subequations}
From the density--density response function, the dielectric function is obtained directly via
\begin{equation}
 \epsilon_\mathrm{gr}^{-1}(q, \omega) = 1+\VC(q)\,\chinn(q,\omega)  \;.
 \label{eq: graphene dielfct}
\end{equation}
As \(\,\mathrm{Im}\,\chinn\,\) is proportional to the dynamic structure factor, \(\mathrm{Im}\,\epsilon_\mathrm{gr}^{-1}\) gives the loss function of graphene. How the individual response functions of Eq.\,\eqref{eq: graphene dielfct} contribute to the Raman spectrum in a 2DEL is nicely explained in \cite{perez2009spin}.
Another major importance of their imaginary part is that they cause a phase delay in the response to an external perturbation, thus giving rise to energy dissipation. 

\begin{figure}[h]
\pgfimage[width=\textwidth]{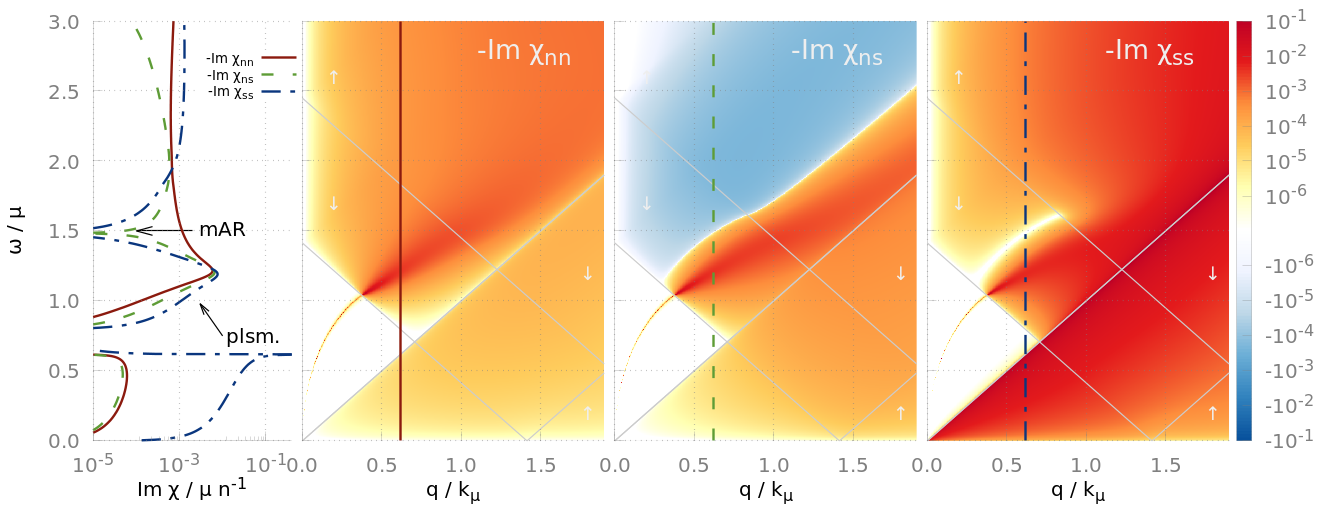}
\caption{Negative imaginary part of the response functions \(\chinn, \chins\) and \(\chiss\)  for spin polarization \(\zeta\!=\!0.5\), in the \((q,\omega)-\)plane (right), as well as at an experimentally accessible wave vector, \(q_0\!\approx 0.62\kmu\) (left). The styles of the vertical \(q_0-\)lines match those of the same function in the left plot. The \(q_0\) plasmon (`plsm.'), having entered the interband PHB, is strongly Landau damped. The magnetic antiresonance (mAR)~\cite{kreil2018resonant} is the prominent curved white region in the rightmost panel; the corresponding dark-blue dash-dotted curve has a distinct gap around \(\omega\!\approx 1.5\mu\), and the green dashed curve becomes zero there.\label{fig:RPA_response_functions} }
\end{figure}

Our results for all three parts of graphene's loss function are shown in Fig.\ref{fig:RPA_response_functions} for a doped system with \(\mu=1800\mathrm{cm}^{-1}\) and a spin-polarization of \(\zeta\!=\!0.5\); (this means that \(n\!\approx 10^{15}\,\mathrm{cm}^{-2}\) and that 75\% of the spins are \(\uparrow\) ).
The plasmon is clearly visible in all three response functions, displaying a \(\sqrt{q}\)-behavior for long wavelengths. The wave vector where it enters the interband PHB is commonly referred to as \(q_\mathrm{c}\), beyond \(q_\mathrm{c}\) the mode gets strongly Landau-damped~\cite{giuliani2005quantum}.

We explicitly point out that both magnetic-field related response functions, \(\chiss\) and \(\chins\), show a distinct lack of excitations above the plasmon in the minority interband PHB (a sign change in \(\chins\) and a white region in \(\chiss\)). Due to the similarity to the 2DEL~\cite{kreil2018resonant, kreil2015excitations}, we call it `magnetic antiresonance' (mAR). It can be understood as follows. An external magnetic perturbation \(B^\mathrm{ext}\) leads to a fluctuation in the magnetization (or spin-density \(s({\bf r})\)), which, for \(\zeta\!\ne\!0\) due to Coulomb coupling implies a fluctuation in the particle-density \(n({\bf r})\) as well:  \(\delta n\!= \chins\, B^\mathrm{ext}\) and \(\delta s\! = \chiss\, B^\mathrm{ext}\). 
The imaginary part of the response functions representing energy absorption, a vanishing of both, \(\,\mathrm{Im}\,\chiss(q,\omega)\) and \(\,\mathrm{Im}\,\chins(q,\omega)\), prohibits magnetic dissipation at these wave vectors and frequencies (the \(\,\mathrm{Re}\) remains finite). This phenomenon is similar to the well known Fano-resonance \cite{fano1935absorption} and fundamental for a binary system (here, \(\uparrow\) and \(\downarrow\)).

%%%%%%%%%%%%%%%%%%%%%%%%%%%%%%%%%%%
\section{SNOM Reflectivity 
 \label{sec: SNOM}}

The basic principle behind s-SNOM is to illuminate the apex of a sharp cantilever above the sample, polarizing the tip. Due to its small curvature the resulting local electric (dipole-)field is very strong. This near-field then interacts with the specimen and is backscattered, sensibly changing, in turn, amplitude and phase of the reflected light far away from the sample. Background scattering from both, tip and surface, is deduced by vibrating the cantilever and demodulation of the detected signal. The method provides a high spatial resolution, probing wavelengths largely independent of the illumination. 

With s-SNOM, density waves can be induced and observed at much larger wave vectors compared to other optical means, \(q\! \gg 1/\lambda_\mathrm{light}\). The accessible \(q\) are in the order of the inverse of the cantilever tip radius \(a\).

The s-SNOM signal strongly depends on the optical properties of the sample, with contributions from the substrate as well as from the graphene sheet.
The reflectivity for {\small\(\mathrm{P}-\)}polarized light of a supporting material with dielectric function \(\kappa_{\scriptscriptstyle\mathrm S}\) is approximated as 
\begin{equation}
   r_{_{\!\mathrm P}} = 1 - \frac{1}{\bar\kappa(\omega)}
  \;,\quad
  \bar\kappa\!= \textstyle\frac12\displaystyle\big(1\!+\! \kappa_{\scriptscriptstyle\mathrm S} \big) \;.
\end{equation}
Here, we use the results measured by Fei et al.~\cite{fei2011infrared}, where we performed a least square fit (see Appendix~\ref{sec:dielectric_sio2} for details). Placing a graphene sheet onto this substrate, changes its reflectivity to
\begin{equation}
 r_{\scriptscriptstyle\mathrm P}= 1 - \frac{1}{\bar\kappa(\omega)-1+\epsilon_\mathrm{gr}(q,\omega)} \;.
\label{equ:rP_gr}\end{equation}
with graphene's dielectric function \(\epsilon_\mathrm{gr}\) given in Eq.\eqref{eq: graphene dielfct}.

The dipole moment \(p_{\scriptscriptstyle\mathrm D}\) induced in the tip is caused by the local field composed of both, the external one, as well as the backscattered field of the sample (typically described by an image dipole). Denoting the polarizability as \(\alpha\), this implies that \(p_{\scriptscriptstyle\mathrm D}\!=\!\alpha E_0 + G\,p_{\scriptscriptstyle\mathrm D}\).
Here, the function \(G\) for the single-dipole approximation gets relevant. 
For a cantilever tip at distance \(d\) it reads~\cite{de2007colloquium, aizpurua2008substrate}
\begin{subequations}\begin{align}
% G_{\tilde{d}}(\omega) &=\, \int\limits_0^\infty \tilde{k}^2 \exp{\!\big(\!-2\,\tilde{k}\,\tilde{d} \big)}\,
 G(\omega;d) &=\, \int\limits_0^\infty \tilde{k}^2 \exp{\!\big(\!-2\,\tilde{k}\tilde{d} \big)}\,
  r_{_{\!\mathrm P\!}}(\tilde{k}, \omega)\,  \mathrm{d}\tilde{k} \\
 &\equiv\> \frac1{I_{\!\tilde{d}}}\, \int\limits_0^\infty g_{\tilde{d}}(\tilde{k})\> 
   r_{_{\!\mathrm P\!}}(\tilde{k}, \omega)\, \mathrm{d}\tilde{k} \;,
\end{align}\label{equ:SNOM_G_FUNCTION}\end{subequations}
with \(I_{\!\scriptscriptstyle\tilde{d}}\!\equiv\!1/4\tilde{d}^3\) and where all lengths are now conveniently measured via the tip-radius: \(\mbox{\small$\tilde{k}$}\!\equiv q\,a\) and \(\mbox{\small$\tilde{d}$}\!\equiv d/a\). 
In Eq.\,\eqref{equ:SNOM_G_FUNCTION} the reflectivity \( r_{\scriptscriptstyle\mathrm P}\) is basically averaged with a distribution function \(g_{\scriptscriptstyle\tilde{d}}(\mbox{\small$\tilde{k}$})\!=\!\mbox{\small$4\tilde{d}^3 \tilde{k}^2$} \exp(-2kd) \) which has its maximum at \(k_{\mathrm mx}\!=\!1/d\). Consequently, \(G(\omega;d)\) is dominated by \( r_{\scriptscriptstyle\mathrm P}(1/d,\,\omega)\).   \;

In s-SNOM, the cantilever is typically operated in tapping mode. To improve the signal to noise ratio, the measured signal \(s\) is proportional to a third order demodulation integral:
\begin{equation}
% s_3(\omega) \sim \int\limits_0^{2\uppi} \frac{\exp(i\,3\,\phi)}{1-G_{\tilde{d}(\phi)}(\omega) } \mathrm{d}\phi
 s_3(\omega) \sim \int\limits_0^{2\uppi} \frac{\exp(i\,3\,\phi)}{1-G\big(\omega;d\mbox{\small$(\phi)$}\big) } \mathrm{d}\phi
 \,=\, 2\uppi\,i\,\sum\limits^{\circlearrowleft}_j \mathrm{Res} f_s(\omega;z_j) \label{equ:SNOM_S}
\end{equation}
with a time periodic distance \(d(\phi)\!=\!d_0 +\, d_1\!\cos(\phi)\). This turns into the sum over all residues in the unit circle of %the function 
\begin{equation}
% f_s(z) \equiv  \frac{-i\,z^2}{1-G_{\tilde{d}(z)}(\omega) }
 f_s(\omega;z) \equiv\,  \frac{-i\,z^2}{1-G(\omega;d\mbox{\small$(z)$}) } \;,
\end{equation}
with the analytically continued distance \({d}(z)\!\equiv\!{d}_0\!+\!\frac{d_1}{2} \left(z\!+\!z^{-1} \right)\). Therefore, the measurement is determined by the poles of \(f_s(z)\), i.e.\ the zeroes of  \(1\!-G(\omega,d\mbox{\small$(z)$})\).
Approximating the function \(g_{\scriptscriptstyle\tilde{d}}(\mbox{\small$\tilde{k}$})\!\approx\!\delta(\mbox{\small$1/\tilde{d}$} - \mbox{\small$\tilde{k}$})\) %\(g_{\scriptscriptstyle\tilde{d}}(\mbox{\small$\tilde{k}$})\!\approx \delta(1\!-\!kd)/\mbox{\small$\tilde{d}$}\)
leads to a better understanding of these singularities. For a free standing graphene sheet, \(\bar\kappa\!=\!1\), the poles are given by \(\, \epsilon_{\mathrm{gr}}(1/d,\omega)\!=\!1/(1-a^{3\!}/4d^3)\,\). This shows that distance and tip radius both significantly influence the measured signal \(s_3(\omega)\), so that recovering the exact plasmon position from the measured signal is highly non-trivial. Much more promising is to evaluate \(s_3(\omega)\) numerically from a model \(r_{\scriptscriptstyle\mathrm P\!}(q,\omega)\) with Eqs.\,\eqref{equ:rP_gr}-\eqref{equ:SNOM_S}, and then compare with the measured data.  

We calculated the reflectivity \(r_{\scriptscriptstyle\mathrm P\!}\) of graphene on SiO\(_2\) for experimentally investigated parameters~\cite{fei2011infrared}. The minority PHB interband edge, lower for higher spin polarization \(\zeta\), causes, the plasmon peak, narrow at \(\zeta\!=0.0\), to get both broader and shifted downwards, until one can no longer distinguish a well-defined collective mode. This picture also nicely demonstrates how  the plasmon and the optical SiO\(_2\) modes to repel each other due to the coupling between graphene and substrate in Eq.\eqref{equ:rP_gr}.

\onecolumngrid  %die leerzeile is wichtig

%\twocolumngrid
%\onecolumngrid  %die leerzeile is wichtig

\begin{figure}[h]
\pgfimage[width=\textwidth]{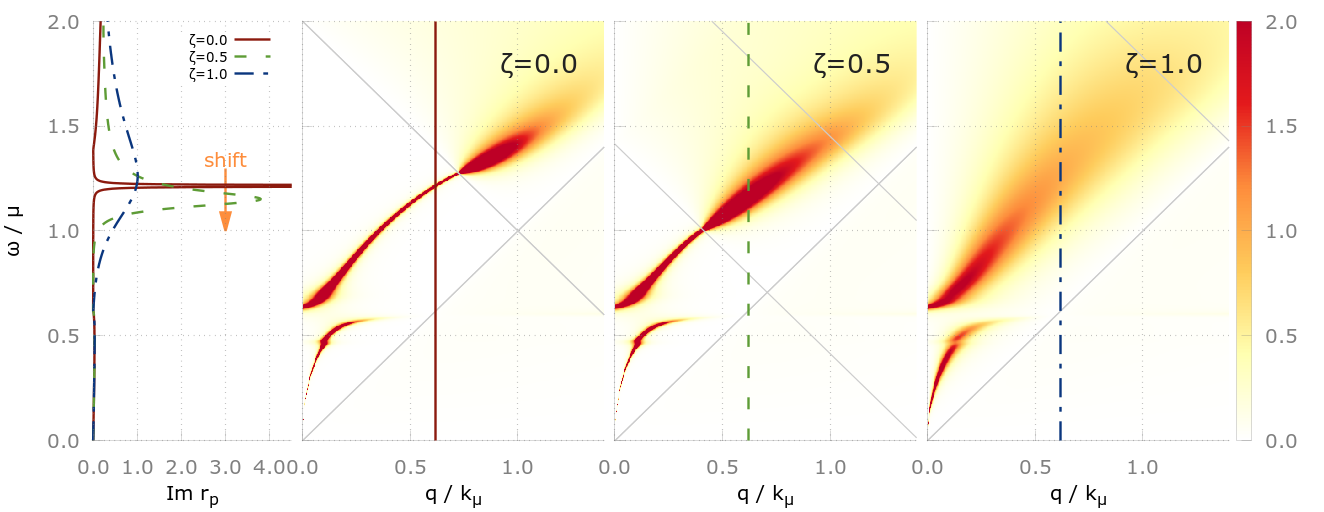}
\caption{Reflectivity \(r_\mathrm{p}\) of graphene on SiO\(_2\) for no, partial, and full polarization (\(\zeta\!=0.0,\, 0.5,\) and \(1.0\), respectively). As in Fig.\,\ref{fig:RPA_response_functions}, the left part shows cuts at \(q_0\!=0.62\kmu\), a typical wave vector for s-SNOM with a tip radius of \(a\!\approx\!30\)nm. The doping level of the graphene sheet corresponds to \(\mu\!=\!1800\) cm\(^{-1}\).  \label{fig_RPA_reflectivity} }
\end{figure}
\twocolumngrid

\ \\

\begin{figure}[h]
\pgfimage[width=0.5\textwidth]{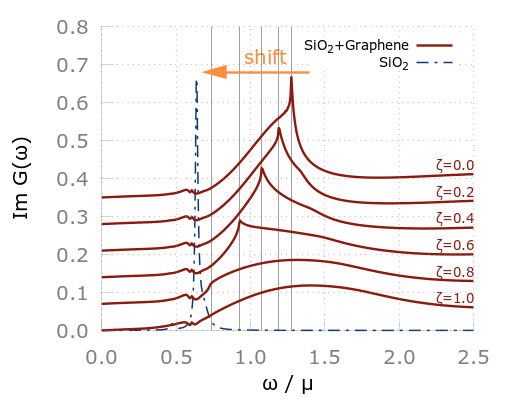}
\caption{Imaginary part of \(G(\omega;d\!=\!0.85a)\), Eq.\,\eqref{equ:SNOM_G_FUNCTION}, for different spin polarizations \(\zeta\) (red solid lines) for graphene on SiO\(_2\). For low \(\zeta\), the plasmon is prominent; larger \(\zeta\) lead to its shift towards lower energies and rigorous damping. 
The narrow (blue dashed) left peak corresponds to an optical mode of the substrate without graphene. All parameters are the same as in Fig.\,\ref{fig_RPA_reflectivity}.
\label{fig:SNOM_FUNCTION_G} }
\end{figure}

For a mean doping level of \(\mu\!=\!1800\,\mathrm{cm}^{-1}\), s-SNOM is sensitive to wave vectors of typically \(0.62\,\kmu\).  Results for different spin polarizations \(\zeta\) are shown in Fig.\,\ref{fig:SNOM_FUNCTION_G} for a  tip radius of \(a\!\approx\!30\mathrm{nm} \).

At a polarization of \(\zeta\!=\!0.6\) the plasmon mode gets strongly damped, as seen in the dipole interaction function \(G(\omega;d)\) in Fig.\ref{fig:SNOM_FUNCTION_G}. Thus, no collective behavior can be observed anymore. 

%%%%%%%%%%%%%%%%%%%%%%%%%%%%%%%%%%%
\section{Conclusions
 \label{sec: Conclusions}}

We have calculated the RPA linear response functions of spin-imbalanced graphene, for a free-standing sheet as well as on a silicon-oxide substrate. 
Similar to the partially spin-polarized 2DEL, graphene also displays a magnetic anti-resonance at \(\omega_{\scriptscriptstyle\mathrm{mAR}}(q)\).
Along this characteristic line in the spectrum, an external magnetic field cannot cause excitations visible in scattering experiments: \(\,\mathrm{Im}\,\chiss\) and \(\mathrm{Im}\,\chins\), describing spin-spin fluctuations and spin-density fluctuations, respectively,  both vanish. 
While \(\mathrm{Im}\,\chins\,\), containing the information on cross-correlations between charge and magnetization just changes sign, \(\,\mathrm{Im}\,\chiss\) is essentially zero in a rather broad \(\omega-\)region over a wide \(q-\)range. 
Interband longitudinal spin excitations (i.e.\ particle-hole excitations of the minority spins with no spin-flip) are drastically suppressed and can no longer contribute to dissipation there.

Compared to the 2DEL~\cite{kreil2018resonant}, graphene's mAR shows several similarities:
First, the intensity of the spin-spin and the density-spin loss function vanish at exactly the same \((q,\omega)\) combinations, underpinning the term mAR. Second, this effect occurs in the PHB of the minority spin electrons, and third it starts exactly where the plasmon starts to be strongly damped: this demonstrates that the coupling between the collective mode and individual particle-hole excitations is responsible for transferring the oscillation energy from one spin species to the other.
However, the mAR in graphene lies at energies above the plasmon, resulting from the fact that interband excitations are the dominating damping mechanism, in contrast to the intraband Landau damping in a 2DEL. 

The conventional plasmon enters the interband PHB at the critical wave vector \(\qc\hspace{0.25pt}\). Increasing the spin polarization \(\zeta\) lowers the minority interband edge and consequently also \(\qc\hspace{0.25pt}\), leading to an earlier Landau-damping. This shifts the collective mode to lower energies and causes an intense broadening. This drastic reduction of lifetime and mean free path of the collective mode opens the door for spin-controlled plasmon transistors: In its working point, the paramagnetic \textit{on}-state has almost no mode damping, whereas in the fully polarized \textit{off}-state the plasmon is thoroughly hindered. 

At finite wave vectors, the dispersion can be observed with s-SNOM, in contrast to other optical setups, restricted to \(q\!\approx\!0\).
In order to test our results experimentally, we therefore applied our theory to the Fresnel coefficient of p-polarized light. This reflectivity \(r_{\mathrm P}(q,\omega)\) is a key quantity entering the dipole interaction function \(G(\omega;d)\) in s-SNOM. We here predict that the plasmon peak is lowered by \(\sim\!30\%\) from the paramagnetic value at \(\zeta\!\approx\!0.6\) and that it will no longer be observable for spin polarizations exceeding this value.

%%%%%%%%%%%%%%%%%%%%%%%%%%%%%%%%%%%
\section*{Acknowledgements}

We thank Zhe Fei for providing us with the measured dielectric function of the substrate. 

%%%%%%%%%%%%%%%%%%%%%%%%%%%%%%%%%%%
\begin{appendix}

\section{Hamiltonian in both Dirac cones
 \label{app: H in Kprime}}

In the close vicinity of the \(K'\) points the pseudospin matrix 
\(\,\underline{\underline{\mbox{\boldmath$\tau$}}}^{\scriptscriptstyle K}_{\,\parallel} \!= 
\big(\underline{\underline{\tau}}_{\,x},\,\underline{\underline{\tau}}_{\,y}\big)\,\)
is replaced by 
\(\underline{\underline{\mbox{\boldmath$\tau$}}}^{\scriptscriptstyle K'}_{\,\parallel} =\big(-\underline{\underline{\tau}}_{\,x},\,\underline{\underline{\tau}}_{\,y}\big)\).
The different sign of the \(x-\)component can be used to define a valley quantum number \(\tau\!= \pm1\)
and thus to write the matrix of the Hamiltonian in Eq.\eqref{equ:DiracH matrix} as
\begin{equation}
   \underline{\underline{\hat{h}}}_{\,\textstyle\tau} 
   =\> \hbar\vF\,\mbox{\small\(
 \begin{pmatrix} 0 &\!\! \frac1{i}\frac{\partial}{\partial x} \!+\! \frac{\partial}{\partial y}\\
   \tau \frac1{i}\frac{\partial}{\partial x} \!-\! \frac{\partial}{\partial y} \!\! &0            
 \end{pmatrix}\)} \;
 \label{equ:DiracH matrix tau}
\end{equation}
in order to treat the two points of the BZ simultaneously.

For better clarity, in addition to their numeric values \(\pm1\), we also use the following labels to distinguish the quantum numbers
\begin{equation*}\begin{array}{llll}
    \sigma \in \{+1, -1\} = \{\,\uparrow,\, \downarrow\,\} &\mbox{spin}\\
    \tau  \in  \{+1, -1\} = \{+, -\} &\mbox{valley} \\
    \ell  \in  \{+1, -1\} = \{\scriptstyle\mathrm C, \scriptstyle\mathrm V\} &\mbox{band} &.
  \end{array}
  \label{equ:label qnumbers}\end{equation*}
\par\medskip

\section{Dielectric function of silicon-oxide 
 \label{sec:dielectric_sio2}}

Using the measured dielectric function of SiO\(_2\) obtained Fei et al.~\cite{fei2011infrared, feiPrivate} we performed a least square fit for these data to the analytic form
\begin{equation}
\epsilon_{\mathrm{S}}(\omega) =\, \epsilon_\infty \!-g_2 \!- g_3 + \sum\limits_{i=1}^3 \frac{g_i \; \omega_i^2}{\omega_i^2-\omega^2-i\;\omega\;\Gamma_i} \label{equ:fitting_epsilon_sio2}
\end{equation}
with \(\epsilon_\infty\!=\!1.85,\, \epsilon_\mathrm{stat}\!=\!2.27\), and 
 \(g_1\! = \epsilon_\mathrm{stat} \!-\! \epsilon_\infty\).
\begin{table}[H]
\centering
\begin{tabular}{r|ccc}
\toprule i: & 1 & 2 & 3 \\ \hline
\(g_i\)      &      & 0.029 & 0.034 \\
\(\omega_i\) & 1072 & 845   & 1237 \\
\(\Gamma_i\) & 29.9 & 49.8  & 147 \\
\bottomrule
\end{tabular}
\caption{Coefficients of Eq.\eqref{equ:fitting_epsilon_sio2} for an SiO\(_2\) bulk substrate (in spectroscopic units: cm\(^{-1}\) for \(\omega_i\) and \(\Gamma_i\))}.
\end{table}

\end{appendix}
%\section*{Bibliography}
%\frenchspacing
%\bigskip

\bibliography{references}

\begin{thebibliography}{43}
\providecommand{\natexlab}[1]{#1}
\providecommand{\url}[1]{\texttt{#1}}
\expandafter\ifx\csname urlstyle\endcsname\relax
  \providecommand{\doi}[1]{doi: #1}\else
  \providecommand{\doi}{doi: \begingroup \urlstyle{rm}\Url}\fi

\bibitem[Mermin(1968)]{mermin1968crystalline}
N.~David Mermin.
\newblock Crystalline order in two dimensions.
\newblock \emph{Phys. Rev.}, 176:\penalty0 250, 1968.

\bibitem[Novoselov et~al.(2004)Novoselov, Geim, Morozov, Jiang, Zhang, Dubonos,
  Grigorieva, and Firsov]{novoselov2004electric}
Kostya~S. Novoselov, Andre~K. Geim, Sergei~V. Morozov, D.~Jiang, Y.~Zhang,
  Sergey~V. Dubonos, Irina~V. Grigorieva, and Alexandr~A. Firsov.
\newblock Electric field effect in atomically thin carbon films.
\newblock \emph{Science}, 306:\penalty0 666--669, 2004.

\bibitem[David et~al.(2004)David, Tsvi, and Steven]{david2004statistical}
Nelson David, Piran Tsvi, and Weinberg Steven.
\newblock \emph{Statistical mechanics of membranes and surfaces}.
\newblock World Scientific, 2004.

\bibitem[Le~Doussal and Radzihovsky(1992)]{le1992self}
Pierre Le~Doussal and Leo Radzihovsky.
\newblock Self-consistent theory of polymerized membranes.
\newblock \emph{Phys. Rev. Lett.}, 69:\penalty0 1209, 1992.

\bibitem[Morozov et~al.(2006)Morozov, Novoselov, Katsnelson, Schedin,
  Ponomarenko, Jiang, and Geim]{morozov2006strong}
Sergei~V. Morozov, Kostya~S. Novoselov, M.I. Katsnelson, F.~Schedin, L.A.
  Ponomarenko, D.~Jiang, and Andre~K. Geim.
\newblock Strong suppression of weak localization in graphene.
\newblock \emph{Phys. Rev. Lett.}, 97:\penalty0 016801, 2006.

\bibitem[Vakil and Engheta(2011)]{VakE11:transformation_optics}
Ashkan Vakil and Nader Engheta.
\newblock Transformation optics using graphene.
\newblock \emph{Science}, 332:\penalty0 1291--1294, 2011.

\bibitem[Yao et~al.(2018)Yao, Liu, Huang, Choi, Xie, Flores, Wu, Yu, Kwong,
  Huang, et~al.]{yao2018broadband}
Baicheng Yao, Yuan Liu, Shu-Wei Huang, Chanyeol Choi, Zhenda Xie, Jaime~Flor
  Flores, Yu~Wu, Mingbin Yu, Dim-Lee Kwong, Yu~Huang, et~al.
\newblock Broadband gate-tunable terahertz plasmons in graphene
  heterostructures.
\newblock \emph{Nature Photonics}, 12:\penalty0 22, 2018.

\bibitem[Garcia~de Abajo(2014)]{garcia2014graphene}
F.~Javier Garcia~de Abajo.
\newblock Graphene plasmonics: challenges and opportunities.
\newblock \emph{ACS Photonics}, 1:\penalty0 135--152, 2014.

\bibitem[Fei et~al.(2011)Fei, Andreev, Bao, Zhang, S.~McLeod, Wang, Stewart,
  Zhao, Dominguez, Thiemens, et~al.]{fei2011infrared}
Zhe Fei, Gregory~O Andreev, Wenzhong Bao, Lingfeng~M Zhang, Alexander
  S.~McLeod, Chen Wang, Margaret~K Stewart, Zeng Zhao, Gerardo Dominguez, Mark
  Thiemens, et~al.
\newblock Infrared nanoscopy of dirac plasmons at the graphene--sio$_2$
  interface.
\newblock \emph{Nano Lett.}, 11:\penalty0 4701--4705, 2011.

\bibitem[Yan et~al.(2013)Yan, Low, Zhu, Wu, Freitag, Li, Guinea, Avouris, and
  Xia]{yan2013damping}
Hugen Yan, Tony Low, Wenjuan Zhu, Yanqing Wu, Marcus Freitag, Xuesong Li,
  Francisco Guinea, Phaedon Avouris, and Fengnian Xia.
\newblock Damping pathways of mid-infrared plasmons in graphene nanostructures.
\newblock \emph{Nature Photonics}, 7:\penalty0 394, 2013.

\bibitem[Low and Avouris(2014)]{low2014graphene}
Tony Low and Phaedon Avouris.
\newblock Graphene plasmonics for terahertz to mid-infrared applications.
\newblock \emph{ACS Nano}, 8:\penalty0 1086--1101, 2014.

\bibitem[Neto et~al.(2009)Neto, Guinea, Peres, Novoselov, and
  Geim]{neto2009electronic}
A.H.~Castro Neto, Francisco Guinea, Nuno~M.R. Peres, Kostya~S. Novoselov, and
  Andre~K. Geim.
\newblock The electronic properties of graphene.
\newblock \emph{Rev. Mod. Phys.}, 81:\penalty0 109, 2009.

\bibitem[Landau(1946)]{landau1946vibrations}
Lev~Davidovich Landau.
\newblock On the vibrations of the electronic plasma.
\newblock \emph{Zh. Eksp. Teor. Fiz.}, 10:\penalty0 25, 1946.

\bibitem[Wolf et~al.(2001)Wolf, Awschalom, Buhrman, Daughton, Von~Molnar,
  Roukes, Chtchelkanova, and Treger]{wolf2001spintronics}
S.A. Wolf, D.D. Awschalom, R.A. Buhrman, J.M. Daughton, S.~Von~Molnar, M.L.
  Roukes, A.~Yu Chtchelkanova, and D.M. Treger.
\newblock Spintronics: a spin-based electronics vision for the future.
\newblock \emph{Science}, 294:\penalty0 1488--1495, 2001.

\bibitem[Agarwal et~al.(2014)Agarwal, Polini, Vignale, and
  Flatt\'e]{APVF14:long-lived}
Amit Agarwal, Marco Polini, Giovanni Vignale, and Michael~E. Flatt\'e.
\newblock Long-lived spin plasmons in a spin-polarized two-dimensional electron
  gas.
\newblock \emph{Phys. Rev. B}, 90:\penalty0 155409, 2014.

\bibitem[Kreil et~al.(2015)Kreil, Hobbiger, Drachta, and
  B{\"o}hm]{kreil2015excitations}
Dominik Kreil, Raphael Hobbiger, J{\"u}rgen~T Drachta, and Helga~M B{\"o}hm.
\newblock Excitations in a spin-polarized two-dimensional electron gas.
\newblock \emph{Phys. Rev. B}, 92:\penalty0 205426, 2015.

\bibitem[Hobbiger et~al.(2017)Hobbiger, Drachta, Kreil, and
  B{\"o}hm]{HDKB17_Phenomenplbroadening}
Raphael Hobbiger, J{\"u}rgen~T. Drachta, Dominik Kreil, and Helga~M. B{\"o}hm.
\newblock Phenomenological plasmon broadening and relation to the dispersion.
\newblock \emph{Solid State Comm.}, 252:\penalty0 54 -- 58, 2017.

\bibitem[Pines(1999)]{pines2018elementary}
David Pines.
\newblock \emph{Elementary Excitations in Solids}.
\newblock Perseus Books, Massachusetts, 1999.

\bibitem[Lee et~al.(2011)Lee, Kim, Bae, Kim, Hong, and Choi]{lee2011optical}
Chul Lee, Joo~Youn Kim, Sukang Bae, Keun~Soo Kim, Byung~Hee Hong, and E.J.
  Choi.
\newblock Optical response of large scale single layer graphene.
\newblock \emph{Appl. Phys. Lett.}, 98:\penalty0 071905, 2011.

\bibitem[Bonitz(1998)]{bonitz1998quantum}
Michael Bonitz.
\newblock \emph{Quantum kinetic theory}.
\newblock Springer, 1998.

\bibitem[Hirjibehedin et~al.(2002)Hirjibehedin, Pinczuk, Dennis, Pfeiffer, and
  West]{hirjibehedin02evidence}
C.~F. Hirjibehedin, A.~Pinczuk, B.~S. Dennis, L.~N. Pfeiffer, and K.~W. West.
\newblock Evidence of electron correlations in plasmon dispersions of ultralow
  density two-dimensional electron systems.
\newblock \emph{Phys. Rev. B}, 65:\penalty0 161309, 2002.

\bibitem[Perez(2009)]{perez2009spin}
Florent Perez.
\newblock Spin-polarized two-dimensional electron gas embedded in a
  semi-magnetic quantum well: Ground state, spin responses, spin excitations,
  and {R}aman spectrum.
\newblock \emph{Phys. Rev. B}, 79:\penalty0 045306, 2009.

\bibitem[Abdelouahed et~al.(2010)Abdelouahed, Ernst, Henk, Maznichenko, and
  Mertig]{AEHM10:spin-split}
Samir Abdelouahed, A.~Ernst, J.~Henk, I.~V. Maznichenko, and I.~Mertig.
\newblock {Spin-split electronic states in graphene: Effects due to lattice
  deformation, {R}ashba effect, and adatoms by first principles}.
\newblock \emph{Phys. Rev. B}, 82:\penalty0 125424, 2010.

\bibitem[Fedorov et~al.(2013)Fedorov, Gradhand, Ostanin, Maznichenko, Ernst,
  Fabian, and Mertig]{FGOM13:impact}
Dmitry~V. Fedorov, Martin Gradhand, Sergey Ostanin, Igor~V. Maznichenko, Arthur
  Ernst, Jaroslav Fabian, and Ingrid Mertig.
\newblock Impact of electron-impurity scattering on the spin relaxation time in
  graphene: A first-principles study.
\newblock \emph{Phys. Rev. Lett.}, 110:\penalty0 156602, 2013.

\bibitem[Dharma-wardana(2006)]{dharma2006coulomb}
M.W.C. Dharma-wardana.
\newblock Coulomb interactions of massless dirac fermions in graphene;
  pair-distribution functions and exchange-driven spin-polarized phases.
\newblock \emph{Solid State Comm.}, 140:\penalty0 4--8, 2006.

\bibitem[Cooper et~al.(2012)Cooper, D’Anjou, Ghattamaneni, Harack, Hilke,
  Horth, Majlis, Massicotte, Vandsburger, Whiteway,
  et~al.]{cooper2012experimental}
Daniel~R. Cooper, Benjamin D’Anjou, Nageswara Ghattamaneni, Benjamin Harack,
  Michael Hilke, Alexandre Horth, Norberto Majlis, Mathieu Massicotte, Leron
  Vandsburger, Eric Whiteway, et~al.
\newblock Experimental review of graphene.
\newblock \emph{ISRN Cond. Matt. Phys.}, 2012:\penalty0 Article {ID} 501686,
  2012.

\bibitem[Wunsch et~al.(2006)Wunsch, Stauber, Sols, and
  Guinea]{wunsch2006dynamical}
B.~Wunsch, T.~Stauber, F.~Sols, and F.~Guinea.
\newblock Dynamical polarization of graphene at finite doping.
\newblock \emph{New J. Phys.}, 8:\penalty0 318, 2006.

\bibitem[Iwamoto(1991)]{Iwam91}
Naoki Iwamoto.
\newblock Static local-field corrections of two-dimensional electron liquids.
\newblock \emph{Phys. Rev. B}, 43:\penalty0 2174--2182, 1991.

\bibitem[Giuliani and Vignale(2005)]{giuliani2005quantum}
Gabriele Giuliani and Giovanni Vignale.
\newblock \emph{Quantum theory of the electron liquid}.
\newblock Cambridge University Press, 2005.

\bibitem[Asgari et~al.(2006)Asgari, Suba{\c{s}}{\i}, Sabouri-Dodaran, and
  Tanatar]{asgari2006static}
R.~Asgari, A.L. Suba{\c{s}}{\i}, A.A. Sabouri-Dodaran, and B.~Tanatar.
\newblock Static local-field factors in a two-dimensional electron liquid.
\newblock \emph{Phys. Rev. B}, 74:\penalty0 155319, 2006.

\bibitem[Moreno and Marinescu(2003)]{moreno2003local}
Juana Moreno and D.C. Marinescu.
\newblock Local-field factors in a polarized two-dimensional electron gas.
\newblock \emph{Phys. Rev. B}, 68\penalty0 (19):\penalty0 195210, 2003.

\bibitem[B{\"{o}}hm et~al.(2010)B{\"{o}}hm, Holler, Krotscheck, and
  Panholzer]{BHKP10:dynamic}
H.~M. B{\"{o}}hm, R.~Holler, E.~Krotscheck, and M.~Panholzer.
\newblock Dynamic many-body theory: Dynamics of strongly correlated fermi
  fluids.
\newblock \emph{Phys. Rev. B}, 82\penalty0 (22):\penalty0 224505, 2010.

\bibitem[Panholzer et~al.(2018)Panholzer, Gatti, and Reining]{PaGR18:nonlocal}
Martin Panholzer, Matteo Gatti, and Lucia Reining.
\newblock Nonlocal and nonadiabatic effects in the charge-density response of
  solids: A time-dependent density-functional approach.
\newblock \emph{Phys. Rev. Lett.}, 120\penalty0 (16):\penalty0 166402, 2018.

\bibitem[Gibertini et~al.(2009)Gibertini, Singha, Pellegrini, Polini, Vignale,
  Pinczuk, Pfeiffer, and West]{GSPP09:engineering}
Marco Gibertini, Achintya Singha, Vittorio Pellegrini, Marco Polini, Giovanni
  Vignale, Aron Pinczuk, Loren~N. Pfeiffer, and Ken~W. West.
\newblock Engineering artificial graphene in a two-dimensional electron gas.
\newblock \emph{Phys. Rev. B}, 79:\penalty0 241406, 2009.

\bibitem[Polini et~al.(2013)Polini, Guinea, Lewenstein, Manoharan, and
  Pellegrini]{PGLM13:artificial}
Marco Polini, Francisco Guinea, Maciej Lewenstein, Hari~C. Manoharan, and
  Vittorio Pellegrini.
\newblock Artificial honeycomb lattices for electrons, atoms and photons.
\newblock \emph{Nature Nanotechnology}, 8:\penalty0 625, 2013.

\bibitem[Wang et~al.(2018)Wang, Scarabelli, Du, Kuznetsova, Pfeiffer, West,
  Gardner, Manfra, Pellegrini, Wind, et~al.]{WSDK18:observation}
Sheng Wang, Diego Scarabelli, Lingjie Du, Yuliya~Y. Kuznetsova, Loren~N.
  Pfeiffer, Ken~W. West, Geoff~C. Gardner, Michael~J. Manfra, Vittorio
  Pellegrini, Shalom~J. Wind, et~al.
\newblock {Observation of Dirac bands in artificial graphene in small-period
  nanopatterned GaAs quantum wells}.
\newblock \emph{Nature Nanotechnology}, 13:\penalty0 29, 2018.

\bibitem[Note1()]{Note1}
Note1.
\newblock The parameter \(r_{\scriptscriptstyle \protect \rm S}\), termed
  \(\alpha \) in graphene, has the constant value of approx. 2.2.

\bibitem[Liu et~al.(2008)Liu, Willis, Emtsev, and Seyller]{liu2008plasmon}
Yu~Liu, Roy~F. Willis, K.V. Emtsev, and Th. Seyller.
\newblock Plasmon dispersion and damping in electrically isolated
  two-dimensional charge sheets.
\newblock \emph{Phys. Rev. B}, 78:\penalty0 201403, 2008.

\bibitem[Dominik et~al.(2018)Dominik, Clemens, Katharina, and
  M.]{kreil2018resonant}
Kreil Dominik, Staudinger Clemens, Astleithner Katharina, and B{\"o}hm~Helga M.
\newblock Resonant and anti-resonant modes of the dilute, spin-inbalanced,
  two-dimensional electron liquid including correlations.
\newblock \emph{Contr. to Plasma Physics}, 58:\penalty0 179--188, 2018.

\bibitem[Fano(1935)]{fano1935absorption}
Ugo Fano.
\newblock Sullo spettro di assorbimento dei gas nobili presso il limite dello
  spettro d’arco.
\newblock \emph{Nuovo Cimento}, 12:\penalty0 154--161, 1935.

\bibitem[De~Abajo(2007)]{de2007colloquium}
F.J.~Garcia De~Abajo.
\newblock Colloquium: Light scattering by particle and hole arrays.
\newblock \emph{Rev. Mod. Phys.}, 79:\penalty0 1267, 2007.

\bibitem[Aizpurua et~al.(2008)Aizpurua, Taubner, de~Abajo, Brehm, and
  Hillenbrand]{aizpurua2008substrate}
Javier Aizpurua, Thomas Taubner, F.~Javier~Garc{\'\i}a de~Abajo, Markus Brehm,
  and Rainer Hillenbrand.
\newblock Substrate-enhanced infrared near-field spectroscopy.
\newblock \emph{Opt. Expr.}, 16:\penalty0 1529--1545, 2008.

\bibitem[Fei()]{feiPrivate}
Zhe Fei.
\newblock Private communication.

\end{thebibliography}

\end{document}